\documentclass{kluwer}
\usepackage{graphicx}

\begin{document}
\begin{article}
\begin{opening}

\title{On the nature of the unidentified X-ray/$\gamma$-ray sources 
IGR~J18027$-$1455 and IGR~J21247+5058}


\author{J.A. \surname{Combi}\email{jcombi@ujaen.es}}
\institute{Universidad de Ja\'en\\
Instituto Argentino de Radioastronom\'{\i}a (IAR)}
\author{M. \surname{Rib\'o}}
\institute{Service d'Astrophysique, CEA Saclay}
\author{I.F. \surname{Mirabel}}
\institute{Service d'Astrophysique, CEA Saclay\\
Instituto de Astronom\'{\i}a y F\'{\i}sica del Espacio (IAFE)}


\runningtitle{On the nature of IGR~J18027$-$1455 and IGR~J21247+5058}
\runningauthor{Combi et~al.}

\begin{ao}
Jorge A. Combi\\
Departamento de F\'{\i}sica\\
Escuela Polit\'ecnica Superior\\
Universidad de Ja\'en\\
Virgen de la Cabeza 2\\
E-23071 Ja\'en\\
Spain\\

\end{ao} 


\begin{abstract} 
We present a multiwavelength study of the environment of the unidentified 
X-ray/$\gamma$-ray sources IGR~J18027$-$1455 and IGR~J21247+5058, recently 
discovered by the IBIS/ISGRI instrument, onboard the INTEGRAL satellite. 
The main properties of the sources found inside their position error circles, 
give us clues about the nature of these high-energy sources.
\end{abstract}

\keywords{X-ray:galaxies, radio continuum:galaxies, infrared:galaxies}



\end{opening}

\section{Introduction}

Since it began to operate in November 2003 until the time of writing (about 18
months), the IBIS/ISGRI instrument onboard the INTEGRAL satellite (Winkler
et~al. \citeyear{winkler03}) has discovered up to 40 unidentified 
X-ray/$\gamma$-ray sources in the energy range from 20 to 100~keV (a regularly
updated list is kept in {\tt http://isdc.unige.ch/$\sim$rodrigue/html/\\
/igrsources.html}). These sources can be high or low mass X-ray binaries (not
detected previously due to the absence of hard X-ray surveys), isolated
pulsars, clusters of galaxies or AGNs, among other objects. New observations
along the electromagnetic spectrum are necessary to discern their nature.

In this work, we present a multiwavelength study of the fields containing the
unidentified INTEGRAL sources IGR~J18027$-$1455 and IGR~J21247+5058 (Walter
et~al. \citeyear{walter04}), as earlier done in Combi et~al.
(\citeyear{combi04}) and in Rib\'o et~al. (\citeyear{ribo04}) but using here
the new positions and information reported in Bird et~al. (\citeyear{bird04}).
Based on the properties of the sources found at other frequencies inside the
error circles in position of the IGR sources, we suggest possible origins for
both high-energy sources.

\section{The IGR~J18027$-$1455 field}

\begin{figure}[t!] 
\center
\resizebox{0.84\hsize}{!}{\includegraphics[angle=0]{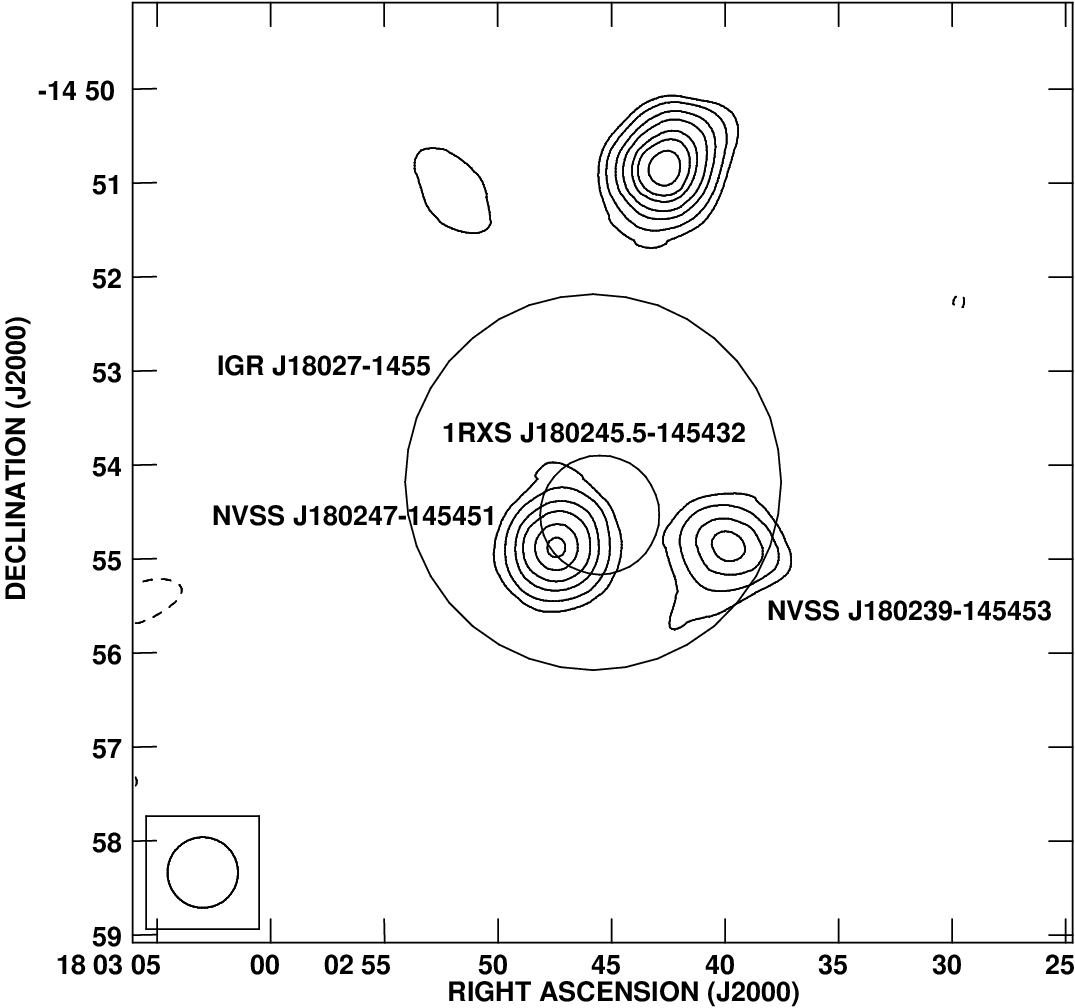}}
\caption{Image of the NVSS data obtained with the VLA at 1.4~GHz on 1997 
October 13 around IGR~J18027$-$1455. The image size is 10$^{\prime}$$\times$10$^{\prime}$. Contours represent $-$3, 3, 5, 8, 11, 15, 18, and 22 times the rms noise level of 0.5~mJy~beam$^{-1}$. The circle in the bottom left corner represents the 45 arcsec of Full Width at Half Maximum (FWHM) of the convolving beam. Two NVSS sources fall inside the 90\% error circle in position of IGR~J18027$-$1455, and one of them is within the 2$\sigma$ uncertainty error circle of a ROSAT source.}
\label{1802a}
\end{figure}

The source IGR~J18027$-$1455 was detected with a significance of 8.9$\sigma$
in the energy range from 20 to 100~keV during 769~ksec of observations
conducted within the INTEGRAL/IBIS survey of the Galactic plane (Bird et~al.
\citeyear{bird04}). The obtained fluxes compared to the Crab are
$F_{(20-40~\rm keV)}$=2.6$\pm$0.2~mCrab and $F_{(40-100~\rm
keV)}$=3.0$\pm$0.4~mCrab. Within its 2 arcmin-radius position error circle (at
90\% confidence or 1.6$\sigma$ uncertainty), centered at $(\alpha,
\delta)_{\rm J2000.0}= (18^{\rm h}02^{\rm m}45{\rlap.}^{\rm s}8,
-14^\circ54^{\prime}11^{\prime\prime})$, we have found two weak point-like
radio sources in the 1.4 GHz NRAO VLA Sky Survey (NVSS, Condon et~al.
\citeyear{condon98}) maps (see Fig.~\ref{1802a}). The first one, the source
NVSS~J180239$-$145453, has an estimated position of $(\alpha, \delta)_{\rm
J2000.0} = (18^{\rm h}02^{\rm m}39{\rlap.}^{\rm s}94 \pm 0{\rlap.}^{\rm s}25,
-14^\circ54^{\prime}53{\rlap.}^{\prime\prime}6 \pm 3{\rlap.}^{\prime
\prime}3)$ (1$\sigma$ uncertainties), has a flux density of 6.9$\pm$0.6~mJy
and has no near infrared (NIR) counterpart in the 2MASS catalog (Cutri et~al.
\citeyear{cutri03}). 

\begin{figure}[t!] 
\center
\resizebox{0.84\hsize}{!}{\includegraphics[angle=0]{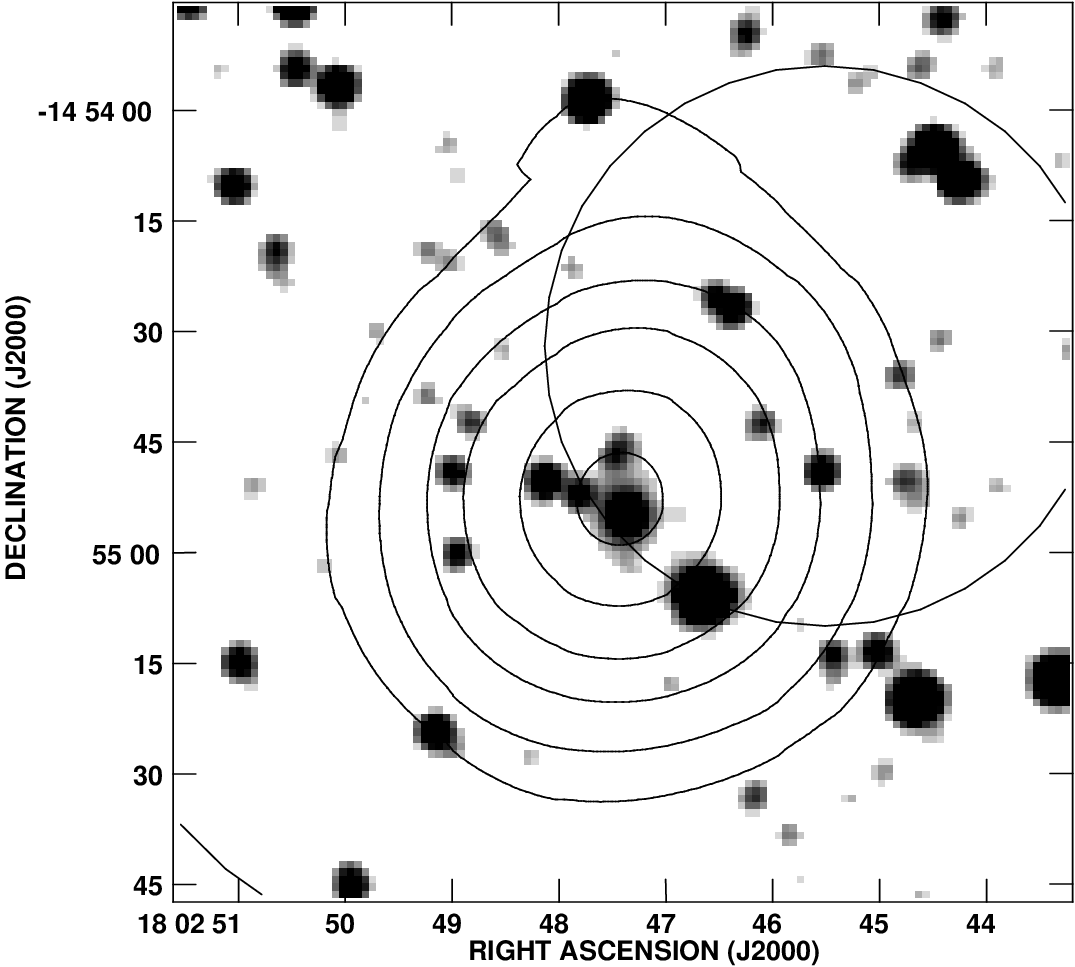}}
\caption[]{Enlargement of a 2$^{\prime}$$\times$2$^{\prime}$ region of Fig.~\ref{1802a} around NVSS~J180247$-$145451, where we plot in greyscale the 2MASS $K_{\rm s}$-band image. The extended NIR source 2MASXi~J1802473$-$145454 is clearly visible in a position compatible with the peak of the radio source.}
\label{1802b}
\end{figure}

The second one, NVSS~J180247$-$145451, has an estimated position of $(\alpha,
\delta)_{\rm J2000.0}=(18^{\rm h}02^{\rm m}47{\rlap.}^{\rm s}37 \pm
0{\rlap.}^{\rm s}13, -14^\circ54^{\prime}51{\rlap.}^{\prime\prime}6 \pm
2{\rlap.}^{\prime\prime}2)$, a flux density of 10.5$\pm$0.6~mJy, and lies
inside and near the edge of the 2$\sigma$ position error circle of the faint
ROSAT X-ray source 1RXS~J180245.5$-$145432 (Voges et~al. \citeyear{voges00}),
which is well within the IBIS/ISGRI error circle. Inside the 2$\sigma$
position error ellipse of this radio source, it is located an extended NIR
source, 2MASXi~J1802473$-$145454 (see Fig.~\ref{1802b}), which has coordinates
$(\alpha, \delta)_{\rm J2000.0}= (18^{\rm h}02^{\rm m}47{\rlap.}^{\rm s}370
\pm 0{\rlap.}^{\rm s}002, -14^\circ54^{\prime}54{\rlap.}^{\prime\prime}76 \pm
0{\rlap.}^{\prime\prime}03)$ and magnitudes $J$=13.18$\pm$0.07,
$H$=12.02$\pm$0.09, $K_{\rm s}$=10.94$\pm$0.04. The optical counterpart of
this NIR source has average magnitudes $B$=19.3$\pm$1.0, $R$=14.9$\pm$0.8 and
$I$=13.8$\pm$0.5 in the USNO-B1.0 catalog (Monet et~al. \citeyear{monet03}).
The photometry of the NIR/optical counterpart is not consistent with a stellar
spectrum. Optical and X-ray observations are planned to confirm the
extragalactic nature of the source and to study its spectrum at softer X-rays.

\section{The IGR~J21247+5058 field}

\begin{figure}[t!] 
\center
\resizebox{0.84\hsize}{!}{\includegraphics[angle=0]{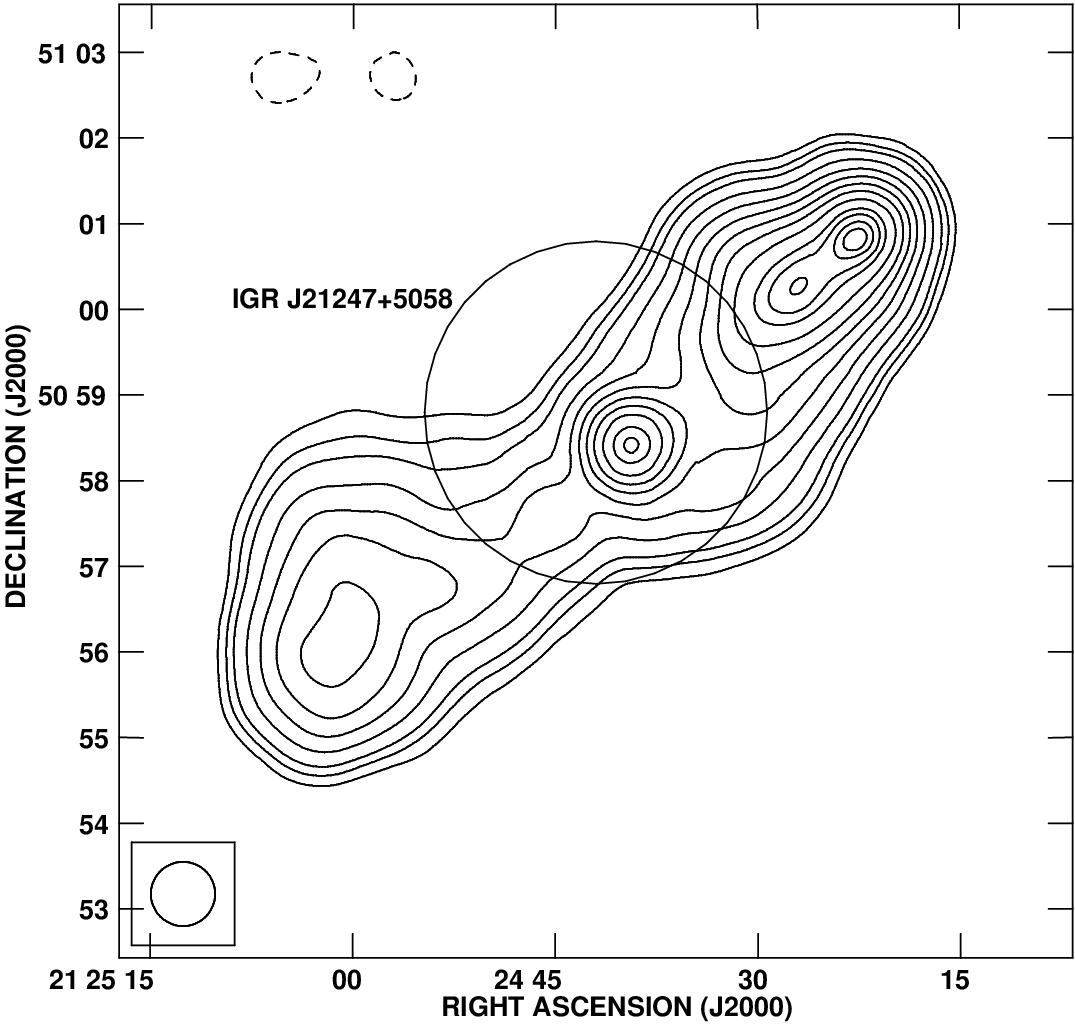}}
\caption[]{Image of the NVSS data obtained with the VLA at 1.4~GHz on 1995 April~3 around IGR~J21247+5058. The image size is 10$^{\prime}$$\times$10$^{\prime}$. The core of the radio source 4C~50.55 is well within the 90\% uncertainty error circle of the IGR source. Contours are $-$3, 3, 6, 10, 20, 35, 55, 80, 110, 150, 200, 230, 260 and 280 times the rms noise level of 1~mJy~beam$^{-1}$. The circle in the bottom left corner represents the 45$^{\prime\prime}$ FWHM of the convolving beam.}
\label{2124a}
\end{figure}

This source was detected with a significance of 6.5$\sigma$ in the energy
range from 20 to 100~keV during only 70~ksec of observations conducted within
the same INTEGRAL/IBIS survey of the Galactic plane (Bird et~al.
\citeyear{bird04}). The obtained fluxes compared to the Crab are
$F_{(20-40~\rm keV)}$=5.4$\pm$0.7~mCrab and $F_{(40-100~\rm
keV)}$=9.3$\pm$1.4~mCrab. Inside its 2 arcmin-radius position error circle (at
90\% confidence or 1.6$\sigma$ uncertainty), centered at $(\alpha,
\delta)_{\rm J2000.0}= (21^{\rm h}24^{\rm m}42{\rlap.}^{\rm
s}0,+50^\circ58^{\prime}48^{\prime\prime})$ is located the core of the bright
radio source 4C~50.55 (see Fig.~\ref{2124a}), also known as
GPSR~93.3194+0.394, KR2, NRAO~659 or BG~2122+50, among other names. 

The core has a flat radio spectrum with a peak flux density of
237~mJy~beam$^{-1}$ at 1.4~GHz. It is at the center of an elongated structure
of 10$\times$3 arcmin (see Fig.~\ref{2124a}) that ends in two large radio
lobes having peak flux densities of 288~mJy~beam$^{-1}$ (NW) and
92~mJy~beam$^{-1}$ (SE), and spectral indices $\alpha~\sim -$0.6 (where
$S_\nu\propto\nu^{+\alpha}$), compatible with optically thin synchrotron
radiation (Mantovani et~al. \citeyear{mantovani82}). The morphology of
4C~50.55 is typical of a radio galaxy, and the comparison among published
radio data shows no significant changes in flux and morphology during the last
30 years (see, e.g. Fanti et~al. \citeyear{fanti81}). These properties are
consistent with an extragalactic source, although, to our knowledge, no
conclusive optical counterpart of a galaxy host has ever been reported,
probably due to its location close to the Galactic plane
($l$=+93.32$^{\circ}$, $b$=+0.39$^{\circ}$). 

\begin{figure}[t!] 
\center
\resizebox{0.84\hsize}{!}{\includegraphics[angle=0]{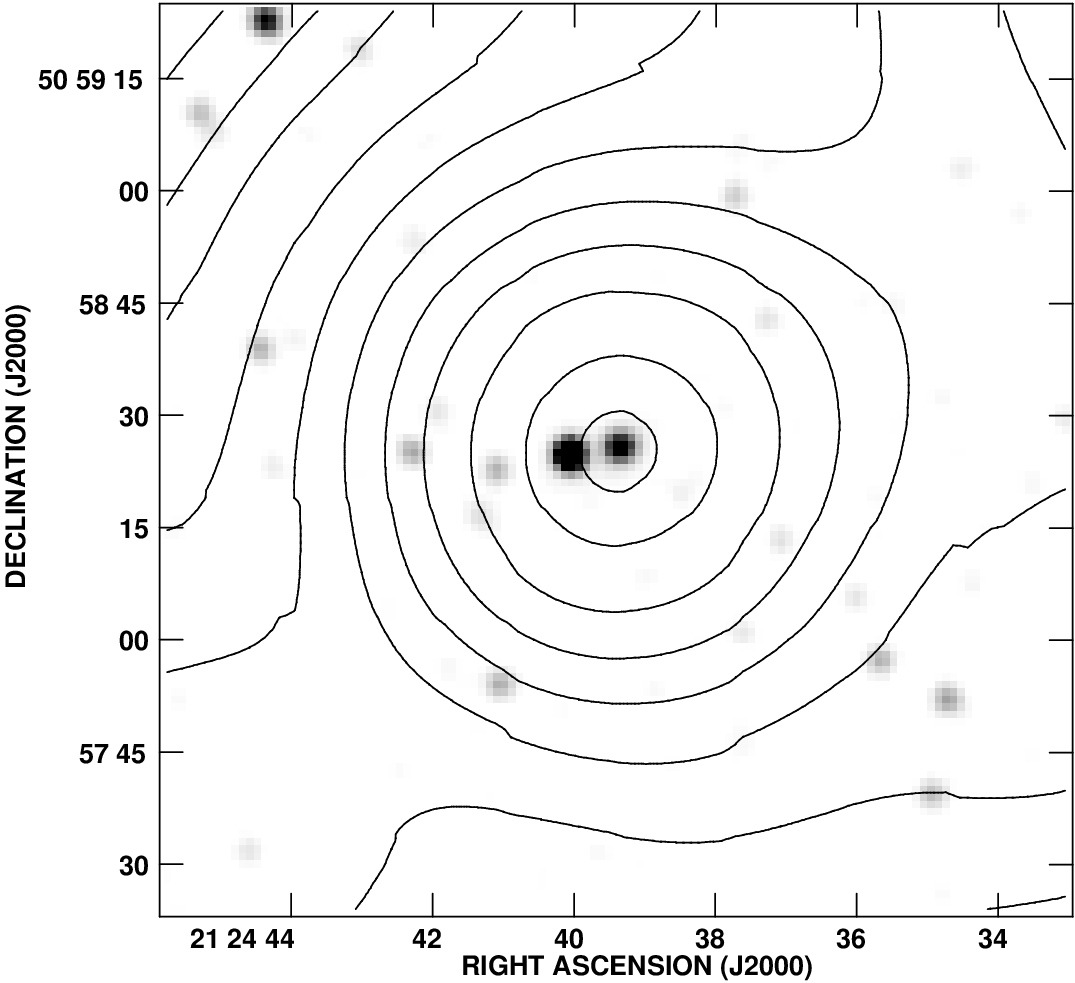}}
\caption[]{Enlargement of a 2$^{\prime}$$\times$2$^{\prime}$ region of Fig.~\ref{2124a} around the core of 4C~50.55, where we plot in greyscale the 2MASS $K_{\rm s}$-band image showing the proposed NIR counterpart, coincident with the radio peak.}
\label{2124b}
\end{figure}

The estimated position of the radio core from the NVSS map and data in the
literature is $(\alpha, \delta)_{\rm J2000.0}=(21^{\rm h}24^{\rm
m}39{\rlap.}^{\rm s}35 \pm 0{\rlap.}^{\rm s}03,
+50^\circ58^{\prime}25{\rlap.}^{\prime\prime}8$\\$ \pm
0{\rlap.}^{\prime\prime}2)$ (68\% or 1$\sigma$ uncertainty). Coincident with
this position there is the NIR source 2MASS~J21243932+5058259 (see
Fig.~\ref{2124b}), with coordinates $(\alpha, \delta)_{\rm J2000.0}= (21^{\rm
h}24^{\rm m}39{\rlap.}^{\rm s}328 \pm 0{\rlap.}^{\rm s}003,
+50^\circ58^{\prime}25{\rlap.}^{\prime\prime}93 \pm
0{\rlap.}^{\prime\prime}03)$ (68\% or 1$\sigma$), and magnitudes
$J$=13.27$\pm$0.04, $H$=12.38$\pm$0.06, $K_{\rm s}$=11.37$\pm$0.04. The
optical counterpart of this NIR source has average magnitudes $B$=16.9$\pm$0.2
and $R$=15.1$\pm$0.2 in the USNO-B1.0 catalog (Monet et~al.
\citeyear{monet03}). The photometry of the NIR/optical counterpart is not
consistent with a stellar spectrum. There is no catalogued ROSAT or EGRET
source coincident with IGR~J21247+5058. Optical and X-ray observations are
planned to confirm the extragalactic nature of the source and to study its
spectrum at softer X-rays.

\section{Further Comments}

We have presented a multiwavelength study of the emission towards the
direction of two recently discovered INTEGRAL sources, namely
IGR~J18027$-$1455 and IGR~J21247+5058. Inside the position error box of the
first one, we have found the radio source NVSS~J180247$-$145451, compatible
with the soft X-ray source 1RXS~J180245.5$-$145432, and having the extended
source 2MASXi~J1802473$-$145454 as NIR counterpart. We suggest that all these
sources are likely associated with the counterpart of the hard X-ray source.
The photometry of the NIR/optical source does not agree with a stellar
spectrum, hence this object seems to be of extragalactic nature. This is
supported by the extended appearance of the NIR counterpart. In the second
case, we have found that the core of an extended radio source, which ends in
two large radio lobes, lies well within the position error circle of the
INTEGRAL source. The morphology of this extended source is typical of a radio
galaxy. Coincident with the position of the radio core is a 2MASS source, with
an optical counterpart not compatible with a stellar spectrum. 

If the X-ray emission of IGR~J21247+5058 comes indeed from a radio galaxy,
then it should originate in the core and/or in the first secction of the jet,
where synchrotron or inverse Compton emission processes dominate (Harris \&
Krawczynski \citeyear{harris02}). This source could be a similar case to
3C~273 (Marshall et~al. \citeyear{marshall01}), where the detected X-rays fade
along the jet. Evidence supporting this possibility was also presented by
Harris et~al. (\citeyear{harrisetal02}) for the radio galaxy M84. In this
object, these authors found an excess of X-ray emission, or knots, aligned
with one of the radio jets. These knots lie at $\sim$ 300 pc from the core and
they are coincident with enhanced radio emission (see their Fig.~1),
suggesting that the X-rays are produced by a synchrotron mechanism.

Recently, Bassani et~al. (\citeyear{bassani04}), have presented a study of
extragalactic sources detected by INTEGRAL during its first year of operation.
Among them, 10 are active galaxies and one is a cluster of galaxies. If the 
extragalactic nature of the sources studied here is confirmed, then many of
the unidentified INTEGRAL sources detected till now could increase
considerably the number of extragalactic objects shining in the hard X-ray
domain.

Finally, it is interesting to note that there is no catalogued EGRET source
coincident with these sources. Multiwavelength observations are currently in
progress to unveil the nature of these high-energy sources.

\acknowledgements

We acknowledge Josep Mart\'{\i} for useful discussions. J.A.C. is a researcher
of the programme {\em Ram\'on y Cajal} funded by the Spanish Ministery of
Science and Technology and the University of Ja\'en, and was also supported by
CONICET (under grant PEI 6384/03).
M.R. acknowledges support by a Marie Curie Fellowship of the European
Community programme Improving Human Potential under contract number
HPMF-CT-2002-02053, and partial support by DGI of the Ministerio de Ciencia y
Tecnolog\'{\i}a (Spain) under grant AYA2001-3092, as well as partial support
by the European Regional Development Fund (ERDF/FEDER).
This research has made use of the NASA's Astrophysics Data System Abstract
Service, of the SIMBAD database, operated at CDS, Strasbourg, France, and of
the NASA/IPAC Extragalactic Database (NED) which is operated by the Jet
Propulsion Laboratory, California Institute of Technology, under contract with
the National Aeronautics and Space Administration. The Digitized Sky Survey
was produced at the Space Telescope Science Institute under U.S. Government
grant NAG~W-2166.
This publication makes use of data products from the Two Micron All Sky
Survey, which is a joint project of the University of Massachusetts and the
Infrared Processing and Analysis Center/California Institute of Technology,
funded by the National Aeronautics and Space Administration and the National
Science Foundation.

\end{article}

\begin{thebibliography}{}

\bibitem[2004]{bassani04}
Bassani, L., Malizia, A., Stephen, J.~B., et~al.
2004, in Proceedings of The 5th INTEGRAL Workshop: The Integral Universe, 
ESA SP-552, in press [{\tt arXiv:astro-ph/0404442}]

\bibitem[2004]{bird04}
Bird, A.~J., Barlow, E.~J., Bassani, L., et~al.
2004, ApJ, 607, L33

\bibitem[2004]{combi04}
Combi, J.~A., Rib\'o, M., \& Mirabel, I.~F.
2004, ATel, 246

\bibitem[1998]{condon98}
Condon, J.~J., Cotton, W.~D., Greisen, E.~W., et~al.
1998, AJ, 115, 1693

\bibitem[2003]{cutri03}
Cutri, R.~M., Skrutskie, M.~F., van Dyk, S., et~al.
2003, VizieR Online Data Catalog, II/246
({\tt http://cdsweb.u-strasbg.fr/viz-bin/Cat?II/246})

\bibitem[1981]{fanti81}
Fanti, C., Mantovani, F., \& Tomasi, P.
1981, A\&AS, 43, 1

\bibitem[2002]{harris02}
Harris, D.~E., \& Krawczynski, H.
2002, ApJ, 565, 244

\bibitem[2002]{harrisetal02}
Harris, D.~E., Finoguenov, A., Bridle, A.~H., Hardcastle, M.~J., \& Laing, R.
2002, ApJ, 580, 110

\bibitem[1982]{mantovani82}
Mantovani, F., Nanni, M., Salter, C.~J., \& Tomasi, P.
1982, A\&A, 105, 176

\bibitem[2001]{marshall01}
Marshall, H.~L., Harris, D.~E., Grimes, J.~P., et~al.
2001, ApJ, 549, L167


\bibitem[2003]{monet03}
Monet, D.~G., Levine, S.~E., Canzian, B., et~al.
2003, AJ, 125, 984
({\tt http://cdsweb.u-strasbg.fr/viz-bin/Cat?I/284})

\bibitem[2004]{ribo04}
Rib\'o, M., Combi, J.~A., \& Mirabel, I.~F.
2004, ATel, 235

\bibitem[2000]{voges00}
Voges, W., Aschenbach, B., Boller, Th., et~al.
2000, VizieR Online Data Catalog, IX/29
({\tt http://cdsweb.u-strasbg.fr/viz-bin/Cat?IX/29})

\bibitem[2004]{walter04}
Walter, R., Bodaghee, A., Barlow, E., et~al.
2004, ATel, 229

\bibitem[2003]{winkler03}
Winkler, C., Courvoisier, T.~J.-L., Di Cocco, G., et~al.
2003, A\&A, 411, L1

\end{thebibliography}
\end{document}